\documentclass[conference]{IEEEtran}

\usepackage{balance}
\usepackage{float}
\usepackage{caption}
\usepackage{subcaption}
\usepackage{tikz,pgfplots,pgfplotstable,booktabs}
\usepackage[official]{eurosym}
\usepackage{xcolor}
\usepackage{colortbl}
\usepackage{comment}
\usepackage{cite}
\usepackage{circuitikz}
\usepackage{amsmath,amsfonts,amsthm,bm}
\usepackage{graphicx}
\usepackage{units}

\usepackage{placeins}
\newcommand{\subparagraph}{}

\usepackage{titlesec}

\allowdisplaybreaks[1]
\usepgfplotslibrary{fillbetween}
\usepgfplotslibrary{statistics}
\usepackage{hyperref}

\pgfplotsset{
    discard if not/.style 2 args={
        x filter/.code={
            \edef\tempa{\thisrow{#1}}
            \edef\tempb{#2}
            \ifx\tempa\tempb
            \else
                
            \fi
    }},
    boxplot/hide outliers/.code={
        \def\pgfplotsplothandlerboxplot@outlier{}}
    }

\usepackage{amsmath,amsfonts,amsthm,amssymb}
\usepackage[mathscr]{euscript}
\usepackage[bb=boondox]{mathalfa}
\usepackage[algoruled]{algorithm2e}
\usepackage{array}
\usepackage{mdframed}

\usepackage[noabbrev,capitalize]{cleveref}

\crefname{equation}{}{}
\Crefname{equation}{}{}

\theoremstyle{definition} 

\theoremstyle{plain} 

\theoremstyle{remark} 

\usepackage{tikz}

\newcommand{\set}[1]{\mathscr{#1}} 
\newcommand{\norm}[1]{\left\lVert#1\right\rVert} 

\makeatletter
\newcommand{\pushright}[1]{\ifmeasuring@#1\else\omit\hfill$\displaystyle#1$\fi\ignorespaces}
\newcommand{\pushleft}[1]{\ifmeasuring@#1\else\omit$\displaystyle#1$\hfill\fi\ignorespaces}
\makeatother

\newcolumntype{C}[1]{>{\centering\arraybackslash}p{#1}} 

\newcounter{box}

\begin{document}


\title{Modeling 100\% Electrified Transportation in NYC\vspace{-0.2cm}}

\author{
\IEEEauthorblockN{
1\textsuperscript{st} Jingrong Zhang, 
2\textsuperscript{nd} Amber Jiang, 
3\textsuperscript{rd} Brian Newborn, 
4\textsuperscript{th} Sara Kou}
\IEEEauthorblockA{\textit{New York University, Center for Urban Science and Progress,}\\ \textit{New York City, USA}}
\and
\IEEEauthorblockN{5\textsuperscript{th} Robert Mieth}
\IEEEauthorblockA{\textit{Princeton University, Dept. of Electrical}\\
\textit{and Computer Engineering, Princeton, USA}}
}


\bstctlcite{IEEE:BSTcontrol}
\maketitle

\begin{abstract}
Envisioning a future 100\% electrified transportation sector, this paper uses socio-economic, demographic, and geographic data to assess electric energy demand from commuter traffic.
We explore the individual mode choices, which allows to create mode-mix scenarios for the entire population, and quantify the electric energy demand for each scenario using technical specifications of battery and electric drives technology in combination with different charging scenarios. 
Using data sets for New York City, our results highlight the need for infrastructure investments, the usefulness of flexible charging policies, and the positive impact of incentivizing micromobility and mass-transit options. 
Our model and results are publicly available as interactive dashboard.
\end{abstract}

\section{Introduction}

Over one quarter of the US greenhouse gas emissions can be traced back to transportation \cite{sourcesofemissions}.
Decarbonizing this sector will impact private households most directly, as drivers must invest in new electric vehicles and adapt their fueling (charging) routines.
Some citizens may instead switch to mass transit or emerging electric micro-mobility such as electric bikes. The exact mode-mixture in a future electrified transportation sector will depend on numerous factors such as availability, personal cost, infrastructure safety, and capacity, leaving significant room for influence from policy-makers and businesses \cite{munoz2016increasing,bueno2017understanding,khan2022inequitable}.
Clearly, due to the varying technical characteristics of different modes of transportation, e.g., battery size, energy efficiency, and charging requirements, the resulting mode-mix will substantially influence the impact of an electrified transportation sector on the power system. 
Yet, comprehensive studies of future charging demands in the context of mode choices are lacking.
This paper proposes a model to quantify mode-mixtures and charging demands as a function of socio-economic data, geographical information, and state-of-the-art electric mobility specifications.  
Studying the case of New York City, USA, we generate an open-source dynamic data set and interactive dashboard to support planning and policy decisions.

The New York State Climate Leadership and Community Protection Act mandates a net-zero carbon economy by 2050 \cite{CLCPA}, implying the abatement of fossil-fueled transportation modes in conjunction with renewable energy generation \cite{dong2014charging}. 
Further, New York State recently passed legislation banning the sale of all gasoline-powered passenger cars and light duty trucks by 2035 \cite{ny2035}. 
With this upcoming ban and ambitious decarbonization goals, New York policy makers are confronted with a significant challenge of understanding the infrastructure needs of this transition. 
With the faded relevance of potential hydrogen-powered vehicles \cite{bakker2010car} a primary concern is the ability of the electric power system to serve the increased demand caused battery-electric vehicle charging facilities \cite{arias2016electric,chadha2022review,acharya2020public} as the share of electricity demand from the transportation sector is expected to rise from \unit[0.2]{\%} to \unit[15--30]{\%} in the next few decades \cite{mai2018electrification}.
While the bulk power system may be able to accommodate this transition, local bottlenecks in dense urban areas and limited distribution system capacity can create barriers for transportation electrification \cite{muratori2021rise}.
To identify such barriers and inform necessary infrastructure investments, estimates of transportation-induced electricity demand is needed. 
Numerous studies, such as \cite{arias2016electric,mangipinto2022impact} and references therein, aim to predict future charging patterns with a focus on local distribution systems or specific charging stations from current car traffic data and contextual information such as weekday or weather. 
Economy-wide assessments of an electrified transportation sector use demographic and socio-economic data into account to estimate energy demands and charging patterns \cite{mai2018electrification}. For example, \cite{mangipinto2022impact} estimates peak demands of around \unit[300]{MW} per \unit[1]{TWh} of yearly energy consumption in Europe. 

\begin{figure*}[h!]
    \centering
    \includegraphics[width=0.8\textwidth]{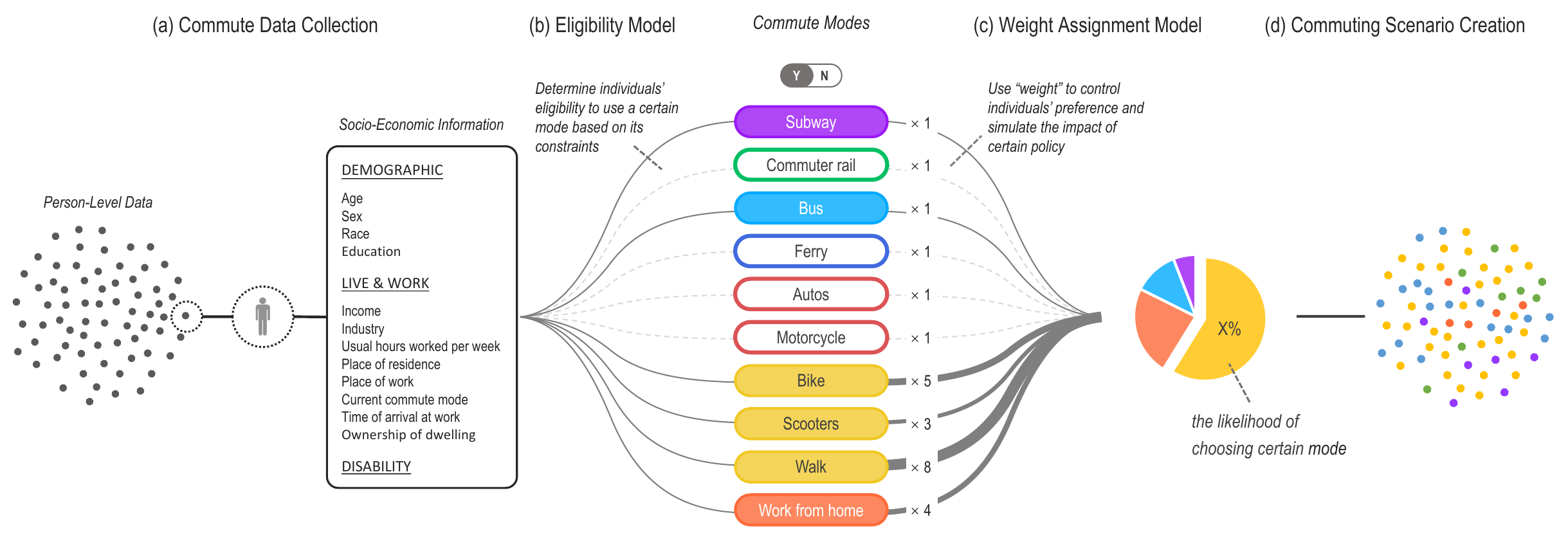}
    \caption{Commuter model: $(a)$ commute data collection, $(b)$ eligibility model, $(c)$ weight assignment model, $(d)$ commuting scenario creation.}
    \label{fig:commuter_model}
    \vspace{-0.5cm}
\end{figure*}

The studies in \cite{chadha2022review,acharya2020public,mai2018electrification,muratori2021rise,arias2016electric,mangipinto2022impact} focus primarily on current car traffic and projected demands from battery-electric vehicles with the same travel patterns.
However, traffic demand management and available infrastructure significantly influence mode choices and, thus, electricity demand patterns. 
For example, a major driver for the dissemination of electric cars is the access to charging infrastructure \cite{gnann2018fast,khan2022inequitable}. 
Micro-mobility (e.g., bicycles) and mass-transit options, on the other hand, are preferred when their infrastructure is perceived as being sufficiently accessible and safe \cite{munoz2016increasing} and can be directly incentivized through public or employer-issued benefits \cite{bueno2017understanding}.
However, to our knowledge, no previous study simultaneously assesses mode-choice and future charging demands.

This paper fills this gap using demographic and socio-economic data from Manhattan, New York, a major urban center that has traditionally relied on transportation modes beyond individual passenger vehicles  (e.g., subway, bicycles).
We focus on modeling individual \textit{commutes} (i.e., regular travel to and from work), given its importance for overall traffic \cite{bueno2017understanding}, and also consider new \textit{work-from-home} trends.
This work uses real-world census data and data from current and emerging electric mobility technology to quantify the transition towards an all-electric transportation sector with regard to power system capacity and individual mode choices. 
Alongside this paper we publish an interactive visualization dashboard (\href{https://tecnyc.herokuapp.com/}{\textbf{tecnyc.herokuapp.com}}), which offers detailed insights for policy- and decision-makers to understand the impact of transportation electrification in New York City.

\section{Commuter Model}

We model the individual mode choices and the resulting aggregate commute scenarios as depicted in Fig.~\ref{fig:commuter_model}.
The model is composed of four parts: $(a)$ commute data collection, $(b)$ eligibility model, $(c)$ weight assignment model, and $(d)$ scenario creation. Sections \ref{ssec:data_description} to \ref{ssec:scenario_creation} below provide details.

\subsection{Commute Data Collection}
\label{ssec:data_description}

Individuals’ commutes and demographics are obtained from the 2019 American Community Survey (ACS) \cite{uscencus} via the IPUMS USA \cite{ruggles2020ipums} online database. The raw data includes demographic information regarding age, gender, race, and education level, residential and work information including income, industry of occupation, usual hours worked per week, place of residence, place of work, typical means of transportation to work, time of arrival at work and ownership of dwelling, as well as disability information. 
Place of residence and work are provided at the the Public Use Microdata Area (PUMA) level. 
In this work, we consider commuters who either work or live in New York County (Manhattan). 
The available data provides the primary means of transportation to work,i.e., the mode used on most days or to cover the greatest distance if the commute requires more than one mode of transportation.
Lastly, we obtain conservative commuting distances as the Cartesian distance between the centroids of each home and place-of-work PUMA using PostGIS.

We categorize commute modes from the ACS data in conjunction with emerging trends like electric scooters to the following groups: transit (subway, commuter rail, bus, ferry), car (autos including private vehicle and taxicab, motorcycle), micro-mobility (bicycle, walk, scooter) and work-from-home (WFH). These categories capture the primary commuting choices in the subsequently derived scenarios.

\subsection{Eligibility Model}

Each commute mode takes relevant individual parameters such as gender, health, and age, as well as contextual parameters, like the existence of safe biking infrastructure, to declare if an individual is eligible for each mode of transportation. These parameters apply different constraints on possible mode choices. 
Taking biking as an example, the maximum age of e-bikers is set to 70 years, maximum travel distance is set to 24 km, ``bikable'' areas include Bronx, Queens, Brooklyn, and Northern NJ where there are available bike infrastructure leading into Manhattan, all genders have equal likelihood to be an eligible bike rider. 
Only individuals who meet all these constraints are eligible for biking.
These constraints can be adapted to tune the model. The details and sources for this implementation can be found in our published model \cite{commmodeldetails}.

As an effect of the COVID-19 pandemic, WFH must be considered as an important factor in commuting pattern estimation.
To predict WFH dynamics in the post-pandemic environment, we build conditional probabilities reflecting any individual's likeliness to WFH at their respective education, profession, and income levels. This data is obtained from the 2020/2021 US Census Household Pulse Survey \cite{hps}. Using this survey's answers about income, education, and WFH status, we are able to find the probability that any American adult works from home given their education status (aggregated to ``not college educated'' and ``college educated'') and their household income (broken down into 7 buckets from \$0--25k to $\ge$\$200k). 
Individuals who work in the service industry or blue-collar professions are less likely to work from home compared to white-collar college educated individuals. 
This calculation is then be used to estimate WFH likelihood in our commuter model.

\subsection{Weight Assignment Model}
Mode weights are then applied to the entire population to generate a scenario. 
Different weights are applied to each commute mode, and modes with higher weights have a higher probability of being utilized. These weights can be adjusted to model the impact of policy decisions, e.g., incentivizing the use of electric bikes or penalizing the use of private cars.

\begin{figure*}[h!]
    \centering
    \includegraphics[width=0.8\textwidth]{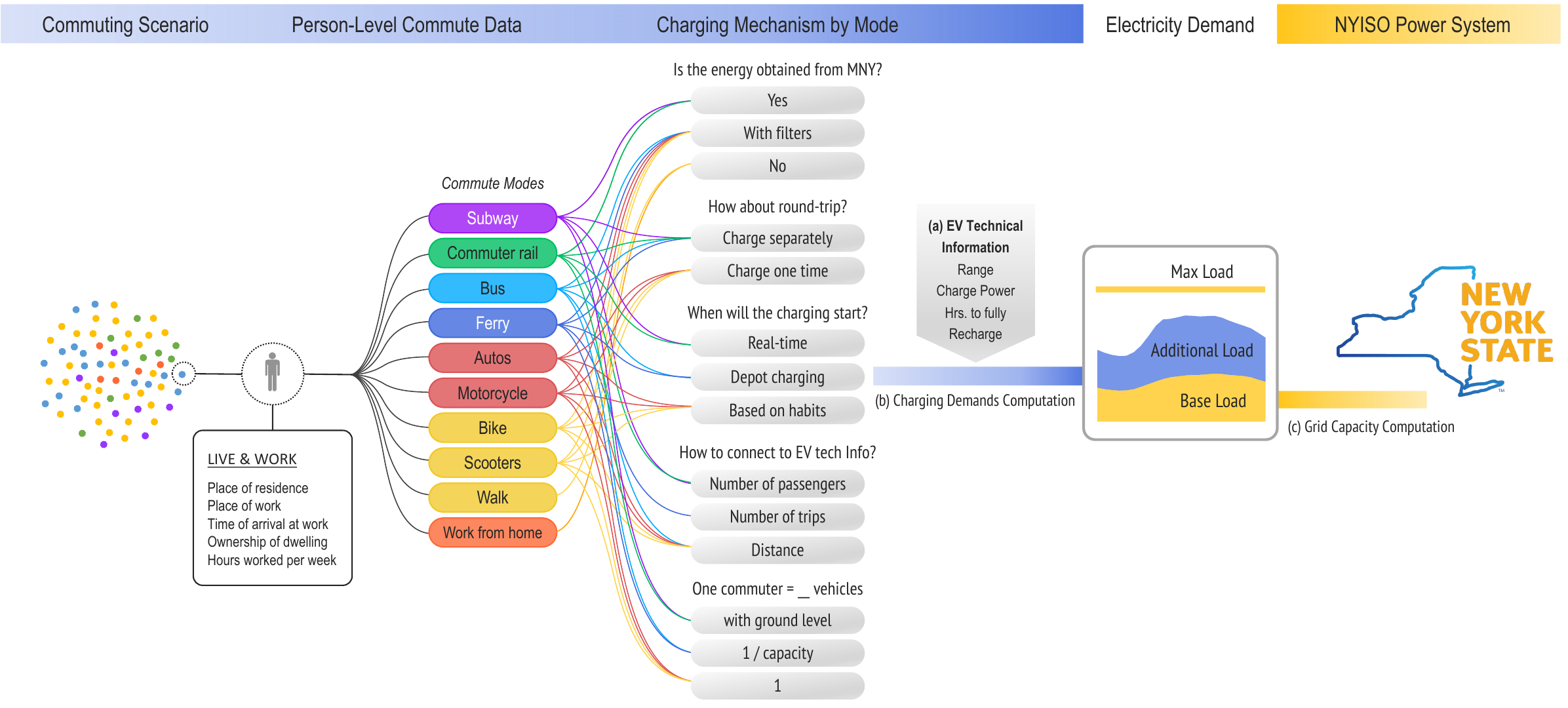}
    \caption{Energy model: $(a)$ EV technical information collection, $(b)$ charging demands computation, $(c)$ grid capacity computation.}
    \label{fig:energy_model}
    \vspace{-0.5cm}
\end{figure*}

\subsection{Commuting Scenario Creation}
\label{ssec:scenario_creation}
We use the commuter model to create four scenarios with various WFH assumptions: A public transit-focused scenario, a car-focused scenario, a micromobility-focused scenario, and a ``Mix'' scenario that represents the most diverse mode choice. 
A baseline 2019 scenario that captures pre-COVID modes of transportation is used as a benchmark. Each of the projection scenarios prioritizes a chosen mode of transportation in the weight assignment model. 
For example, if in the public transit-focused scenario an individual has more than one choice of commuting methods, the model is more likely to assign that individual to take transit over other options.
Note that our model supports the creation of customized scenarios beyond those presented here. 
See also our description of the dashboard and model publication in Section~\ref{sec:Dashboard} below.

\section{Energy Model}
\label{sec:energy_model}

Any commuting scenario will result in its own unique energy requirements given mode choices and travel distances.
We propose a framework for comparing the electricity demand identified in the commuter model with the current state of the New York Independent System Operator (NYISO) power system as illustrated in Fig.~\ref{fig:energy_model}. 
The Energy Model has three parts: $(a)$ EV technical information collection, $(b)$ charging demand computation, and $(c)$ grid capacity computation. 
Sections \ref{ssec:ev_technical_information} to \ref{ssec:computing_grid_capacity} below provide details.

\subsection{EV Technical Information Collection}
\label{ssec:ev_technical_information}
We collect the technical characteristics of available electric modes of transportation including range (km), efficiency (kWh/km), and charging power (kW) from EV databases and manufacturers, e.g., \cite{evdatabase,newflyer}. 
For subway and commuter rail, which requires power at the time of motion, we estimate the average hourly fixed electricity demand and the average additional load per individual based on the annual power usage, hourly ridership, and operating lines in New York City.

\subsection{Energy Demands Computation}
\label{ssec:copmuting_charging_demands}

We obtain the per-mode hourly energy demand by combining a given input commuting scenario, i.e., a mode assignment for each trip, with the parameters of our electric mobility efficiency and battery capacity database.
This considers the difference between transportation modes that require electricity at the time of use, like subways or commuter rails, and modes that require post-use charging, like EVs or e-bikes.

Current EV charging technology typically provides charging power immediately upon arrival until the battery is sufficiently charged.
However, flexible (smart) charging, e.g., driven by price incentives or directly controlled by the power utility, is envisioned to play an important role in future grid operations \cite{dong2014charging}. 
We model these options through three charging scenarios:``Earliest'' reflects the current state-of-technology of charging upon arrival, ``Latest'' assumes just-in-time charging so that each vehicle is charged briefly before it is scheduled for its next use, ``Distributed'' minimizes load peaks by distributing charging start times of each vehicle as evenly as possible throughout the day, while respecting the energy demand and vehicle schedule.

Further, we consider charging location based on mode type and commuter demographics.
Only bus commuters from within Manhattan will impact charging demands as other boroughs or states will charge their buses upon returning after a trip. Additionally, we assume that personal EVs will only be charged once (either at work or at home) to cover a full round-trip commute. 
Here, single-family house owners are modeled more likely to charge at home compared to apartment dwellers.

\subsection{Grid Capacity Computation}
\label{ssec:computing_grid_capacity}

Generation capacity and transmission infrastructure set an upper bound for the maximum instantaneous power that can be supplied to Manhattan. 
We quantify this bound as a critical reference for transportation electricity demand as follows.
We use a synthetic but realistic network model of the New York power system with 1,814 buses, 2,202 high voltage transmission lines, and 395 generators \cite{liang2022weather}.
For given power injections from all generators at time $t$, collected in vector $\bm{p}_t$, and power withdrawals from loads at all buses at time $t$, collected in vector $\bm{d}_t$, we approximate the resulting vector of power flows $\bm{f}_t$ using a linear mapping (DC power flow model) $\bm{f}_t =\bm{B}(\bm{M}\bm{p}_t - \bm{d}_t)$. Here, matrix $\bm{B}$ collects the so called power transfer distribution factors and $\bm{M}$ maps generators to buses.  
Denoting $\overline{\bm{p}}$, $\underline{\bm{p}}$ as the vectors collecting upper and lower production limits of each generator and $\overline{\bm{f}}$ as the vector collecting the maximum flow capacity of each transmission line we can find the maximum power that can be served inside Manhattan by solving the following program:
\begin{subequations}
\begin{align}
\max_{\bm{a}_t \ge 0, \bm{p}_t, \bm{x}} \  
    &\sum_{t\in\set{T}}\Big[ \bm{e}^T \bm{a}_t - \lambda \norm{\bm{a}_t \circ \bm{d}_t^{-1}}_2^2\Big] \label{pf_model:objective} \\
\text{s.t.}\ \forall t \in \set{T}: \quad
    & a_{i,t} = 0,\ \forall i\notin\set{M} \label{pf_model:non_manhattan_buses} \\
    & \bm{e}^T \bm{p}_t = \bm{e}^T (\bm{d}_t + \bm{a}_t) \label{pf_model:balance}\\
    & \underline{\bm{p}} \le \bm{p} \le \overline{\bm{p}} \label{pf_model:gen_limits} \\
    & -\overline{\bm{f}} \le \bm{B}(\bm{M}\bm{p}_t -  \bm{d}_t - \bm{a}_t) \le \overline{\bm{f}}  \label{pf_model:flow_limits} \\
    & \bm{p}_t \in \set{X}(\bm{x}, \bm{d}_t) \label{pf_model:aux_constraints}
\end{align}%
\label{eq:pf_model}%
\end{subequations}%
Objective \cref{pf_model:objective} maximizes the additional load $\bm{a}_t$ indexed as $[a_{i,t}]$ for all timesteps $t\in\set{T}$, which we set to be 24 hours of one day. The second term in \cref{pf_model:objective} is a regularization term that ensures that the relative change of load is similar across all buses. Operator $\circ$ is the element-wise product, $\bm{e}$ is the vector of ones, and $\norm{\cdot}_2$ the 2-norm.
Constraint \cref{pf_model:non_manhattan_buses} defines the buses to which additional load can be added via set $\set{M}$, e.g., all buses within Manhattan. Constraint \cref{pf_model:balance} ensures that the system is balanced and constraints \cref{pf_model:gen_limits,pf_model:flow_limits} enforce generation and transmission capacities. Lastly, constraint \cref{pf_model:aux_constraints} captures additional operational constraints, such as reserve requirements. The full set of constraints follows \cite[Eq.~(5)]{liang2022weather}.

We use real-world demand data published by the New York Independent System Operator \cite{nyiso_data} for demand profiles $\bm{d}_t$ and solve problem \cref{eq:pf_model} for two summer days (Aug.~3 and 28) and two winter days (Feb.~14 and 21) in 2018. These days have been chosen to combine high and low demand peaks with high and low wind power generation, respectively.
Set $\set{M}$ collects the indices of all buses located in Manhattan south of 125th~St.

\begin{figure*}
    \centering
    \includegraphics[width=\textwidth]{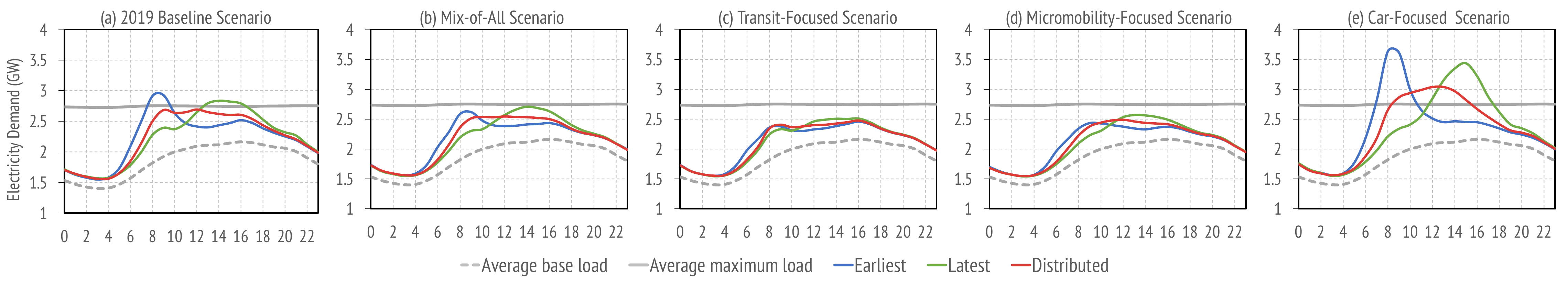}
    \caption{Electricity demand of (a) 2019 baseline and (b)--(e) various policy scenarios in three different charging start times. We adapt medium work-from-home level to policy scenarios as this level is closest to the 2019 baseline scenario.}
    \label{fig:energy_result}
    \vspace{-0.5cm}
\end{figure*}

The resulting average maximum power that can be delivered to the buses in $\set{M}$ is approximately \unit[2700]{MW} as shown in Fig.~\ref{fig:energy_result}.
This limit is mainly driven by the limit of transmission capacity as we observed no significant sensitivity to the amount of wind generation or demand outside of NYC. 
Note that this limit is an optimistic limit as it assumes no demand increase outside of the buses $\set{M}$.

\begin{figure}
    \centering
    \includegraphics[width=\linewidth]{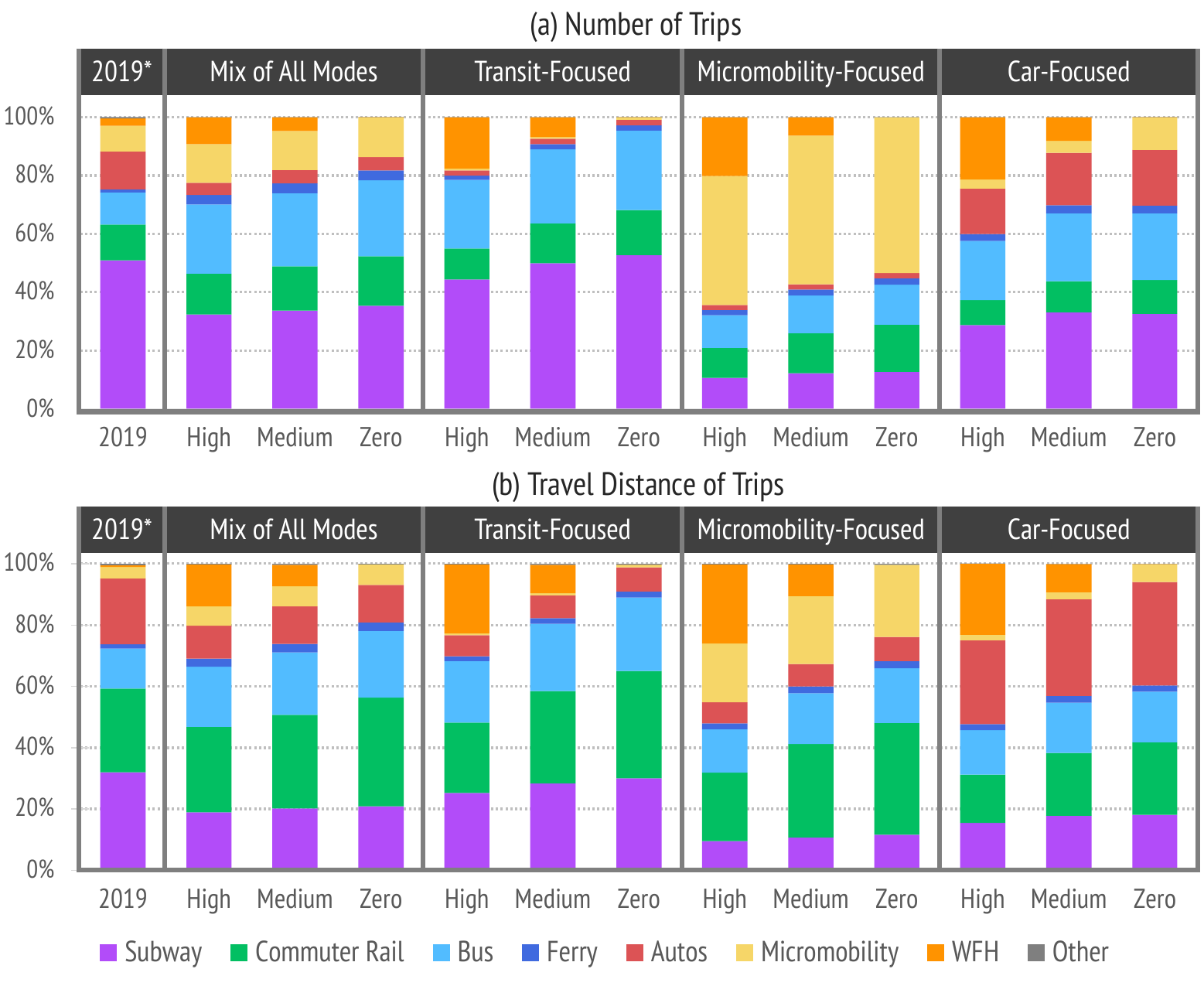}
    \caption{The share of commute modes on (a) number of trips and (b) travel distance in each scenario. $x$ axis is the WFH Level (high, medium and zero).}
    \label{fig:scenarios}
\end{figure}

\section{Results and Dashboard}

\subsection{Mode Scenarios}

The studied scenarios obtained using the method described in Section~\ref{ssec:scenario_creation} are shown in Fig.~\ref{fig:scenarios} as the share of each mode on the number of trips and travel distance. Compared to the 2019 baseline scenario, there is potential to favor micro-mobility given our eligibility criteria. Alternatively, there is less potential for additional capacity for public transit, particularly subway ridership, because its usage is already approaching limits in current reachable areas and incentivizing public transit requires cost new infrastructure, such as new lines and stations. 
Notably, the increase in the share of car users has a greater impact on the share of travel distance, as there is no maximum distance limit for cars.

\subsection{Energy Demand}

Fig.~\ref{fig:energy_result} shows the computed electricity demand as described in Section~\ref{sec:energy_model} relative to the estimated power system limit (Section~\ref{ssec:computing_grid_capacity}). 
For most scenarios, charging peaks will be close to or higher than the computed system limit.
Scenarios with lower demand are characterized by high participation in public transit and micromobility and  represent large shift in mode choice relative to the 2019 base-point. 
Even for the more moderate ``Mix'' scenario, demand is close to the system limit.

Demand strongly depends on the underlying mode mix and on the charging policy. (``Earliest'', ``Latest'', or ''Distributed'' as introduced in Section~\ref{ssec:copmuting_charging_demands} above.)
``Earliest'', reflecting current state-of-the-art technology, creates a morning peak after commuters arrive at work. 
``Latest'' creates a flatter peak in the afternoon, because different energy demands (resulting from different travel distances) and different working hours lead to more time-distributed charging.
Finally, ``Distributed'' charging, which we use to approximate a peak-minimizing charging pattern, shows the flattest charging profile.
In the 2019 baseline scenario (Fig.~\ref{fig:energy_result}(a)), there is an over \unit[200]{MW} difference in demand between the scenarios of drivers choosing to charge at the start of work vs. randomly distributed.
Notably, charging is mainly driven by electric cars, leading to most extreme peaks in the ``Car-Focused'' scenario (Fig.~\ref{fig:energy_result}(e)).  

The results shown in Fig.~\ref{fig:energy_result} allow three central conclusions. 
First, investments to increase power system capacity will be needed to accommodate the energy demands from an electrified transportation sector. Even in the scenarios with the lowest estimated peak demand (Fig.~\ref{fig:energy_result}(c)), about 92\% of the (optimistically) computed power system limit is reached.
Second, policies that incentivize commuters to prioritize micromobility or mass transit over cars will directly reduce resulting energy demands without impacting commuter mobility. 
This will both reduce necessary infrastructure investment cost and reduce electricity prices through lower peak demands.
(Note that the scenarios leading to the results in Fig.~\ref{fig:energy_result} only prioritize certain modes within the current space of possible modes for each commuter. Polices that change the mode options for each commuter, e.g., by creating additional bicycle infrastructure or expanding mass transit access may have even more significant impacts on mode choices.)
Third, our results highlight the value of incentivizing smart charging and demand response technology allowing time-distributed charging.

\begin{figure}[b]
    \centering
    \includegraphics[width=0.99\linewidth]{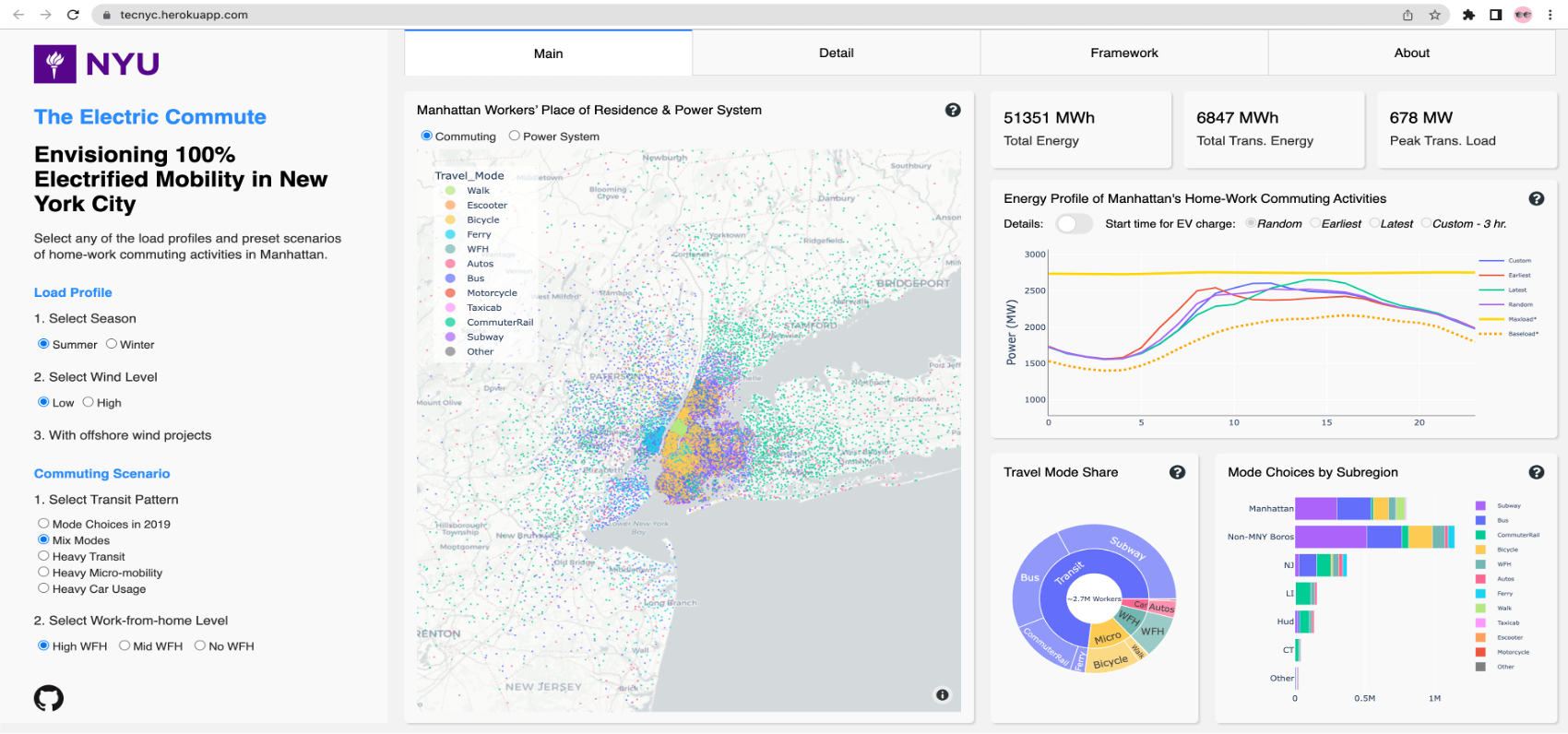}
    \caption{The landing page of the dashboard.}
    \label{fig:dashboard}
\end{figure}

\subsection{Dashboard}
\label{sec:Dashboard}
This work is published as an open-source model \cite{code_git} and interactive dashboard (\href{https://tecnyc.herokuapp.com/}{\textbf{tecnyc.herokuapp.com}}) to aid users to explore, analyze, and understand the models and their resulting data in greater detail. 
The landing page of the dashboard consists of functional areas including the control sidebar, map panel, energy profile, and commute profile, as shown in Fig.~\ref{fig:dashboard}. 
The dashboard provides both an overview and a detailed breakdown to further help users identify patterns and gain insights. It can be utilized by a wide range of users from interested citizens to policy makers who seek to quantify impacts of potential policy decisions.

\section{Conclusion}
Fossil fuel-based transportation modes must be electrified to decarbonize the transportation sector, challenging existing electric power infrastructure. 
Focusing on New York City, we used 2019 ACS commuting survey responses to create a baseline commuting pattern and created a commuter model that captures the dynamics of individual mode choices with individual parameters. From this we derived four scenarios with an emphasis on transit, micromobility, car, and a mix of modes. 
Using battery and electric drive data we computed the electric energy requirements of each scenario and compared it to an estimation of the maximum power system capacity.
The Car-Focused scenario exceeds the maximum grid capacity, which informs policy makers and infrastructure planners that either the grid should be heavily upgraded or policy should be passed to incentivize individuals to shift to other modes of transportation. Policy makers and city planners can utilize these models to help inform future electrical planning. 
Possible future work includes extensions to stochastic- or agent-based models with context information (e.g., weather) as in \cite{mangipinto2022impact}.

\bibliographystyle{IEEEtran}
\bibliography{literature}

\begin{thebibliography}{10}
\providecommand{\url}[1]{#1}
\csname url@samestyle\endcsname
\providecommand{\newblock}{\relax}
\providecommand{\bibinfo}[2]{#2}
\providecommand{\BIBentrySTDinterwordspacing}{\spaceskip=0pt\relax}
\providecommand{\BIBentryALTinterwordstretchfactor}{4}
\providecommand{\BIBentryALTinterwordspacing}{\spaceskip=\fontdimen2\font plus
\BIBentryALTinterwordstretchfactor\fontdimen3\font minus
  \fontdimen4\font\relax}
\providecommand{\BIBforeignlanguage}[2]{{%
\expandafter\ifx\csname l@#1\endcsname\relax
\typeout{** WARNING: IEEEtran.bst: No hyphenation pattern has been}%
\typeout{** loaded for the language `#1'. Using the pattern for}%
\typeout{** the default language instead.}%
\else
\language=\csname l@#1\endcsname
\fi
#2}}
\providecommand{\BIBdecl}{\relax}
\BIBdecl

\bibitem{sourcesofemissions}
\BIBentryALTinterwordspacing
{Sources of Greenhouse Gas Emissions}. [Online]. Available:
  \url{https://www.epa.gov/ghgemissions/sources-greenhouse-gas-emissions}
\BIBentrySTDinterwordspacing

\bibitem{munoz2016increasing}
B.~Mu{\~n}oz \emph{et~al.}, ``The increasing role of latent variables in
  modelling bicycle mode choice,'' \emph{Transp. Rev.}, vol.~36, no.~6, pp.
  737--771, 2016.

\bibitem{bueno2017understanding}
P.~C. Bueno \emph{et~al.}, ``Understanding the effects of transit benefits on
  employees’ travel behavior,'' \emph{Transportation Research Part A: Policy
  and Practice}, vol.~99, pp. 1--13, 2017.

\bibitem{khan2022inequitable}
H.~A.~U. Khan \emph{et~al.}, ``{Inequitable access to EV charging
  infrastructure},'' \emph{The Electricity Journal}, vol.~35, no.~3, p. 107096,
  2022.

\bibitem{CLCPA}
\BIBentryALTinterwordspacing
{Climate Leadership and Community Protection Act (Climate Act)}. [Online].
  Available: \url{https://climate.ny.gov/Our-Progress}
\BIBentrySTDinterwordspacing

\bibitem{dong2014charging}
J.~Dong \emph{et~al.}, ``{Charging infrastructure planning for promoting
  battery electric vehicles},'' \emph{Transportation Research Part C: Emerging
  Technologies}, vol.~38, pp. 44--55, 2014.

\bibitem{ny2035}
\BIBentryALTinterwordspacing
{Governor Hochul Drives Forward New York's Transition to Clean Transportation}.
  [Online]. Available:
  \url{https://www.governor.ny.gov/news/governor-hochul-drives-forward-new-yorks-transition-clean-transportation}
\BIBentrySTDinterwordspacing

\bibitem{bakker2010car}
S.~Bakker, ``The car industry and the blow-out of the hydrogen hype,''
  \emph{Energy Policy}, vol.~38, no.~11, pp. 6540--6544, 2010.

\bibitem{arias2016electric}
M.~B. Arias \emph{et~al.}, ``Electric vehicle charging demand forecasting model
  based on big data technologies,'' \emph{Applied Energy}, vol. 183, 2016.

\bibitem{chadha2022review}
S.~Chadha \emph{et~al.}, ``{A review on Smart Charging impacts of Electric
  Vehicles on Grid},'' \emph{Materials Today: Proceedings}, 2022.

\bibitem{acharya2020public}
S.~Acharya \emph{et~al.}, ``{Public plug-in electric vehicles+ grid data},''
  \emph{IEEE Trans. Smart Grid}, vol.~11, no.~6, pp. 5099--5113, 2020.

\bibitem{mai2018electrification}
T.~T. Mai \emph{et~al.}, ``Electrification futures study,'' National Renewable
  Energy Lab.(NREL), Golden, CO (United States), Tech. Rep., 2018.

\bibitem{muratori2021rise}
M.~Muratori \emph{et~al.}, ``The rise of electric vehicles—2020 status and
  future expectations,'' \emph{Progress in Energy}, vol.~3, no.~2, p. 022002,
  2021.

\bibitem{mangipinto2022impact}
A.~Mangipinto \emph{et~al.}, ``{Impact of mass-scale deployment of electric
  vehicles and benefits of smart charging across all European countries},''
  \emph{Applied Energy}, vol. 312, p. 118676, 2022.

\bibitem{gnann2018fast}
T.~Gnann \emph{et~al.}, ``{Fast charging infrastructure for electric vehicles:
  Today’s situation and future needs},'' \emph{Transportation Research Part
  D: Transport and Environment}, vol.~62, pp. 314--329, 2018.

\bibitem{uscencus}
\BIBentryALTinterwordspacing
{United States Census Data}. [Online]. Available: \url{https://data.census.gov}
\BIBentrySTDinterwordspacing

\bibitem{ruggles2020ipums}
S.~Ruggles \emph{et~al.}, ``{IPUMS USA: Version 10.0. Minneapolis, MN},'' 2020.

\bibitem{commmodeldetails}
\url{https://github.com/BNewborn/mobility-electrification/blob/main/05_Commuter_Electric_Pipeline/commuter_model/}.

\bibitem{hps}
\BIBentryALTinterwordspacing
{Household Pulse Survey}. [Online]. Available:
  \url{https://www.census.gov/programs-surveys/household-pulse-survey/datasets.html}
\BIBentrySTDinterwordspacing

\bibitem{evdatabase}
\BIBentryALTinterwordspacing
{Electric Vehicle Database}. [Online]. Available:
  \url{https://ev-database.org/}
\BIBentrySTDinterwordspacing

\bibitem{newflyer}
\BIBentryALTinterwordspacing
{New Flyer’s battery-electric, zero-emission bus}. [Online]. Available:
  \url{https://www.newflyer.com/bus/xcelsior-charge-ng/}
\BIBentrySTDinterwordspacing

\bibitem{liang2022weather}
Z.~Liang \emph{et~al.}, ``{Weather-Driven Flexibility Reserve Procurement},''
  \emph{arXiv preprint arXiv:2209.00707}, 2022.

\bibitem{nyiso_data}
\BIBentryALTinterwordspacing
NYISO. {Energy Market \& Operational Data}. [Online]. Available:
  \url{www.nyiso.com/energy-market-operational-data}
\BIBentrySTDinterwordspacing

\bibitem{code_git}
\BIBentryALTinterwordspacing
{Code Supplement - Electrified Mobility}. [Online]. Available:
  \url{https://github.com/BNewborn/mobility-electrification}
\BIBentrySTDinterwordspacing

\end{thebibliography}

\end{document}